\documentclass[twocolumn,footinbib,superscriptaddress,showpacs]{revtex4}%
\usepackage{graphicx}
\usepackage[latin1]{inputenc}
\usepackage{amsmath}
\usepackage{amsfonts}
\usepackage{amssymb}%
\providecommand{\U}[1]{\protect\rule{.1in}{.1in}}
\begin{document}
\title{Measurement of the quadratic Zeeman shift of $^{85}$Rb hyperfine sublevels
using stimulated Raman transitions}
\author{Run-Bing Li}
\author{Lin Zhou}
\affiliation{State Key Laboratory of Magnetic Resonance and Atomic and Molecular Physics,
Wuhan Institute of Physics and Mathematics, Chinese Academy of Sciences, Wuhan
430071, China}
\affiliation{Center for Cold Atom Physics, Chinese Academy of Sciences, Wuhan 430071, China}
\affiliation{Graduate School, Chinese Academy of Sciences, Beijing 100080, China}
\author{Jin Wang}
\email{wangjin@wipm.ac.cn}
\author{Ming-Sheng Zhan}
\affiliation{State Key Laboratory of Magnetic Resonance and Atomic and Molecular Physics,
Wuhan Institute of Physics and Mathematics, Chinese Academy of Sciences, Wuhan
430071, China}
\affiliation{Center for Cold Atom Physics, Chinese Academy of Sciences, Wuhan 430071, China}


\begin{abstract}
\textbf{Abstract}

We demonstrate a technique for directly measuring the quadratic Zeeman shift
using stimulated Raman transitions. The quadratic Zeeman shift has been
measured yielding $\Delta\nu=1296.8$ $\pm3.3$ Hz/G$^{2}$ for magnetically
insensitive sublevels ($5S_{1/2},F=2,m_{F}=0\rightarrow$ $5S_{1/2}%
,F=3,m_{F}=0$) of $^{85}$Rb by compensating the magnetic field and cancelling
the ac Stark shift. We also measured the cancellation ratio of the
differential ac Stark shift due to the imbalanced Raman beams by using two
pairs of Raman beams ($\sigma^{+}$, $\sigma^{+})$ and it is $1$:$3.67$ when
the one-photon detuning is $1.5$ GHz in the experiment.

\end{abstract}
\maketitle

\begin{flushleft}
\textbf{ 1. Introduction}
\end{flushleft}

Since the atom interferometer was demonstrated in 1991\cite{Kasevich1991a}, it
has been applied to rotation measurement, such as inertial navigation and even
the rotation rate of the earth \cite{Gustavson2000a,Canuel2006a}. Recently, an
atom-interferometer gyroscope of high sensitivity and long-term stability was
reported \cite{Durfee2006a}. In order to improve the accuracy of the rotation
rate measurement by using an atom-interferometer gyroscope, the potential
systematic errors should be considered and controlled as well as possible. The
quadratic Zeeman shift is considered as a factor that influences the accuracy
of the rotation rate measurement in the atom-interferometer gyroscope.

The atom gyroscope generally uses two counter-propagating cold-atom clouds
launched in strongly curved parabolic trajectories \cite{Canuel2006a}. The two
cold atom clouds should be overlapped completely in order to cancel common
noise and gravity acceleration, and cold collisions occur between atoms along
similar trajectories. For a dual atom-interferometer gyroscope, Rubidium is a
suitable candidate because it has a smaller collision frequency shift than
Cesium \cite{Santos2002a,Kokkelmans1997a,Gibble1995a,Sortais2000a}. In our
previous work \cite{Li2008a,Wang2007a}, we have experimentally investigated
the stimulated Raman transitions in the cold atom interferometer. Both the
accuracy and the fringe contrast of an atom-interferometer gyroscope can be
improved by studying the magnetic field dependence of the coherent population
transfer. A homogenous magnetic field must be applied along the Raman beams to
keep the quantization axis consistent and resolve degenerate magnetic
sublevels. This magnetic field will cause Zeeman shifts. The quadratic Zeeman
shift induces a relative frequency shift of the two coherent states, which
influences the accuracy of the rotation rate measurement. It is therefore
important to measure accurately and understand the quadratic Zeeman shift of
$^{85}$Rb in the cold atom interferometer. Similarly, the quadratic Zeeman
shift is important in other applications such as microwave frequency standards
\cite{Ramsey1963a,Thomas1982a,Hemmer1986a}, optical frequency standards
\cite{Kajita2005a,Boyd2007a} and coherent population trapping clock
\cite{Vanier2005a}. The quadratic Zeeman shift can be usually obtained from
the Breit-Rabi formula after the magnetic field is measured by the linear
Zeeman effect \cite{Bize1999a}. We study this from the field-insensitive clock
transitions whose linear Zeeman shift is zero, thus the magnetic field is
calibrated from other $u_{F}\neq0$ states. We have also studied this quadratic
Zeeman shift in the presence of the ac Stark shift of the Raman pulses.

In this paper, we analyze the hyperfine sublevels of the ground states in the
magnetic field by using second-order perturbation theory, and demonstrate
experimentally the coherent population transfer of the different Zeeman
sublevels by stimulated Raman transitions. The quadratic Zeeman shift of the
ground state of $^{85}$Rb was measured by the two-photon resonance of the
stimulated Raman transition after the ac Stark shift was cancelled and the
residual magnetic field was compensated. The value of the magnetic field is
calibrated by the linear Zeeman shift. Our analysis shows that the quadratic
Zeeman shift can be measured to Hz level for magnetically insensitive states
($5S_{1/2},F=2,m_{F}=0\rightarrow$ $5S_{1/2},F=3,m_{F}=0$) in our experiment.
We also measured the cancellation ratio of the differential ac Stark shift due
to the imbalanced Raman beams by using two pairs of Raman beams. This study
provides useful data for higher precision measurement of the quadratic Zeeman
shift of $^{85}$Rb, even for improving the accuracy of the rotation rate
measurement of the atom-interferometer gyroscope.

\begin{flushleft}
\textbf{2. Quadratic Zeeman shift}
\end{flushleft}

Including the hyperfine interaction, the ground state energy levels will split
and shift in the magnetic field. The interaction Hamiltonian operator
\cite{Sobelman1996a,Itano2000a} within the subspace of hyperfine sublevels
associated with the electronic levels is given by%

\begin{equation}
H^{^{\prime}}=hA_{S}I\cdot J+g_{J}\mu_{B}J\cdot B+g_{I}\mu_{B}I\cdot B
\label{h-1}%
\end{equation}

where, $h$ is the Plank constant, $A_{S}$\ is the hyperfine constant, $I$ and
$J$ are the nuclear spin operators and orbital angular momentum respectively,
$g_{J}$ and $g_{I}$ are the electronic $g$-factor and nuclear $g$-factor
respectively, $\mu_{B}$ is Bohr magneton. Second-order perturbation theory is
valid for low magnetic-field intensity, and the energies of the hyperfine
Zeeman sublevels for the ground states can be derived as following

For F=2%

\begin{equation}
E(\frac{1}{2},2,0,B)=E(\frac{1}{2})-\frac{7}{4}hA_{S}-\mathstrut
\smallskip\frac{(g_{J}-g_{I})^{2}}{12hA_{S}}\mu_{B}^{2}B^{2} \label{Q-1}%
\end{equation}

\begin{align}
E(\frac{1}{2},2,\pm1,B)  &  =E(\frac{1}{2})-\frac{7}{4}hA_{S}\mp\frac
{g_{J}-7g_{I}}{6hA_{S}}\mu_{B}B\label{Q-2}\\
&  -\frac{2(g_{J}-g_{I})^{2}}{27hA_{S}}\mu_{B}^{2}B^{2}\nonumber
\end{align}

\begin{align}
E(\frac{1}{2},2,\pm2,B)  &  =E(\frac{1}{2})-\frac{7}{4}hA_{S}\mp\frac
{g_{J}-7g_{I}}{3hA_{S}}\mu_{B}B\label{Q-3}\\
&  -\frac{5(g_{J}-g_{I})^{2}}{108hA_{S}}\mu_{B}^{2}B^{2}\nonumber
\end{align}

For F=3%

\begin{equation}
E(\frac{1}{2},3,0,B)=E(\frac{1}{2})+\frac{5}{4}hA_{S}+\frac{(g_{J}-g_{I})^{2}%
}{12hA_{S}}\mu_{B}^{2}B^{2} \label{Q-4}%
\end{equation}

\begin{align}
E(\frac{1}{2},3,\pm1,B)  &  =E(\frac{1}{2})+\frac{5}{4}hA_{S}\pm\frac
{g_{J}+5g_{I}}{6hA_{S}}\mu_{B}B\label{Q-5}\\
&  +\frac{2(g_{J}-g_{I})^{2}}{27hA_{S}}\mu_{B}^{2}B^{2}\nonumber
\end{align}

\begin{align}
E(\frac{1}{2},3,\pm2,B)  &  =E(\frac{1}{2})+\frac{5}{4}hA_{S}\pm\frac
{g_{J}+5g_{I}}{3hA_{S}}\mu_{B}B\label{Q-6}\\
&  +\frac{5(g_{J}-g_{I})^{2}}{108hA_{S}}\mu_{B}^{2}B^{2}\nonumber
\end{align}

\begin{equation}
E(\frac{1}{2},3,\pm3,B)=E(\frac{1}{2})-\frac{7}{4}hA_{S}\pm\frac{g_{J}+5g_{I}%
}{2hA_{S}}\mu_{B}B \label{Q-7}%
\end{equation}

Here, $E(J,F,m_{F},B)$\ denotes the energy of the hyperfine sublevels,
including the effect of the hyperfine interaction and magnetic field
splitting. From eqs.(\ref{Q-1}) and (\ref{Q-4}), the quadratic Zeeman shift
for the transition $5S_{1/2},F=2,m_{F}=0\rightarrow5S_{1/2},F=3,m_{F}=0$ is
$\Delta\nu=(g_{J}-g_{I})^{2}\mu_{B}^{2}B^{2}/6hA_{S}$, which is consistent
with the reference \cite{Steck} that is obtained from the Breit-Rabi formula
when it is extended to second order in the field strength.

\begin{flushleft}
\textbf{3. Experimental configuration}
\end{flushleft}

%

\begin{figure}
[ptbh]
\begin{center}
\includegraphics[
trim=1.023722in 3.617887in 1.202633in 3.733542in,
height=2.0401in,
width=3.0346in
]%
{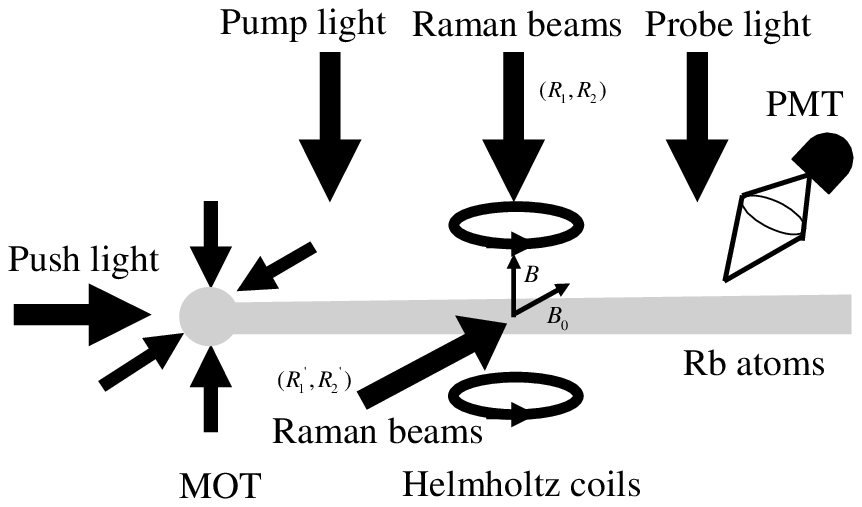}%
\caption{Experimental scheme: cold $^{85}$Rb atoms fly horizontally from the
MOT to the probe region. Three crossed pairs of Helmholtz coils are applied to
compensate the residual magnetic field in the stimulated Raman interaction
area. The combined Raman beams ($R_{1}$,$R_{2}$) and ($R_{1}^{^{\prime}}%
$,$R_{2}^{^{\prime}}$) are parallel to the magnetic field $B$ and $B_{0}$
respectively. The laser-induced fluorescence signal is detected by a PMT. }%
\label{f-1}%
\end{center}
\end{figure}

The experimental arrangement is shown in Fig.\textit{\ref{f-1}}, which is
similar to our previous work \cite{Li2008a,Wang2007a}. Briefly, the cold atoms
are trapped in a nonmagnetic stainless steel chamber with $14$ windows, where
the trapping and repumping beams are provided by a tapered amplifier diode
laser (TOPTICA TA100) and an external-cavity diode laser (TOPTIC DL100)
respectively, whose frequencies are stabilized using saturated absorption
spectroscopy \cite{Wang2000a}. After the polarization gradient cooling (PGC)
process, the atoms are guided by a near-resonance laser pulse and fly
transversely from the trapping region to the probe region at a velocity of
$24$ m/s \cite{Jiang2005a}. Then, they are completely pumped to the ground
state $5S_{1/2},F=2$ as the initial state by a perpendicular linearly
polarized laser beam which is near resonance with the transition
$5S_{1/2},F=3\rightarrow5P_{3/2},F=2$. Three crossed pairs of Helmholtz coils
are used to provide the magnetic field in the Raman interaction area, where
the current of the coils along the Raman beams (R$_{1}$, R$_{2}$) is
controlled by the DC power supply (MPS-901) and measured by the digital
multimeters (Flucke 8846A). The magnetic field intensity is calibrated by the
first-order Zeeman shift, whose uncertainty is less than one part in one
thousand. The combined Raman beams (R$_{1}$, R$_{2}$) and (R$_{1}^{^{\prime}}%
$, R$_{2}^{^{\prime}}$) are applied along the magnetic fields $B$ and $B_{0}$
respectively in the stimulated Raman interaction region.\textit{ }The Raman
beams (R$_{1}$, R$_{2}$) are used to measure the frequency shift induced by
the external fields such as the Raman beams (R$_{1}^{^{\prime}}$,
R$_{2}^{^{\prime}}$) and the magnetic field $B$. The Raman beams (R$_{1}$,
R$_{2}$) and (R$_{1}^{^{\prime}}$, R$_{2}^{^{\prime}}$) are supplied from the
same Raman laser. This configuration has the benefit for the accurate
measurement of the ac stark shift because two pairs of Raman beams always have
the same one-photon detuning. The detailed description of the Raman laser
arrangement is similar to our previous work \cite{Wang2007a}. The atoms are
transferred to the state $5S_{1/2},F=3$ from $5S_{1/2},F=2$ when they pass
through a Raman $\pi$-pulse. After coherent population transfer via a
simulated Raman transition, the population of the state is detected by a laser
induced fluorescence (LIF) signal, and we use a photo multiplier tube (PMT) to
collect the LIF.%

\begin{figure}
[ptbh]
\begin{center}
\includegraphics[
trim=1.277246in 3.101367in 1.729737in 4.200657in,
height=2.1508in,
width=2.7397in
]%
{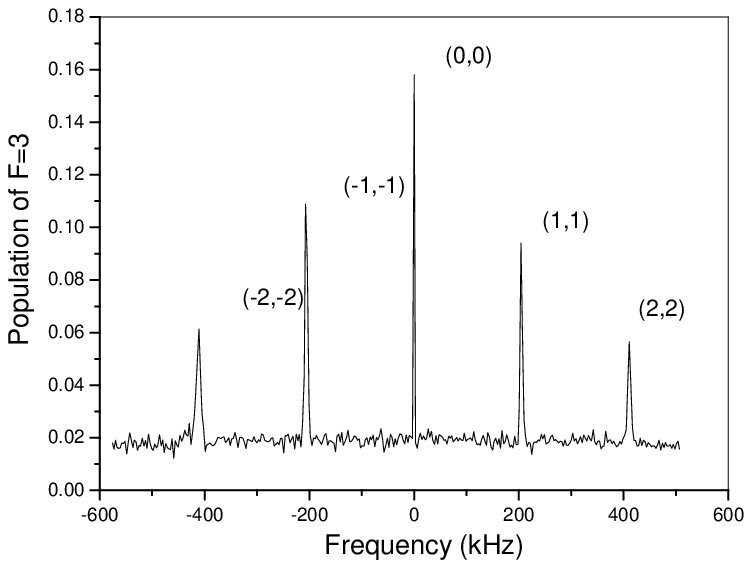}%
\caption{Population transfer dependence on two-photon detuning. The peaks
(-2,-2),(-1,-1),(0,0),(1,1),(2,2) are the resonance transition of the
hyperfine Zeeman sublevels when the Raman beams ($R_{1}$,$R_{2}$) are applied
along the magnetic field $B=220$ mG, where the Raman beams ($R_{1}^{^{\prime}%
}$,$R_{2}^{^{\prime}}$) and $B_{0}$ are not used.}%
\label{f-2}%
\end{center}
\end{figure}

\begin{flushleft}
\textbf{4. Results and analysis}
\end{flushleft}%

\begin{figure}
[ptbh]
\begin{center}
\includegraphics[
trim=1.018908in 3.268674in 1.570081in 3.927799in,
height=2.0392in,
width=2.738in
]%
{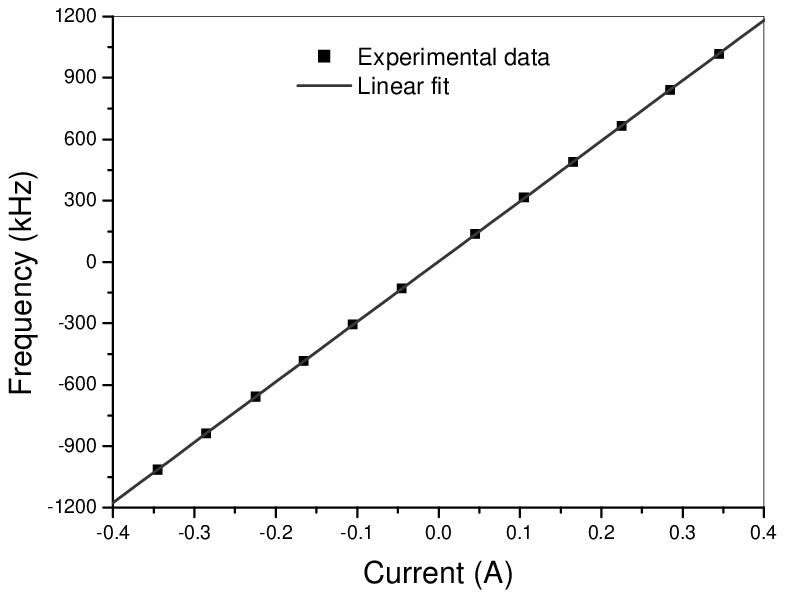}%
\caption{The resonance frequency of the hyperfine Zeeman sublevels (-2,-2) and
(2,2) depends on the current of the Helmholtz coils. It is measured by Raman
beams (R$_{1}$, R$_{2}$) under R$_{1}^{^{\prime}}$=$0$, R$_{2}^{^{\prime}}%
$=$0$, B$_{0}$=$0$. The magnetic field can be scaled by the first Zeeman shift
after the linear fit, whose uncertainty is below one per thousand.}%
\label{f-3}%
\end{center}
\end{figure}

The hyperfine level $F=2$ is split into five sublevels and $F=3$ into seven
sublevels, whose energies are expressed as in eqs.(\ref{Q-1}-\ref{Q-7}) when
there exists a magnetic field. After the magnetic field is compensated
completely according to our previous work \cite{Li2008a}, all sublevels are
degenerate. Coherent population transfer can occur for the transition of the
combined hyperfine Zeeman sublevels $(-2,-2),(-1,-1),(0,0),(1,1),(2,2)$ when
the Raman beams (R$_{1}$, R$_{2}$) with ($\sigma^{+}$,$\sigma^{+}$) propagate
along the magnetic field $B$ in the stimulated Raman interaction region as
shown in Fig.\textit{\ref{f-1}} \cite{Li2008a,Peters,Gustavson,Petelsk}, where
the Raman beams (R$_{1}^{^{\prime}}$, R$_{2}^{^{\prime}}$) and the magnetic
field $B_{0}$ are not used. The maximum population transfer is achieved when
two-photon resonance is satisfied with the transition selection rules shown in
Fig.\textit{\ref{f-2}}. A perfect symmetric Raman spectrum are achieved when
the atoms are interacted with Raman beams ($\sigma^{+}$,$\sigma^{+}$). In
Fig.\textit{\ref{f-2}, }the transition probability can be explained using the
oscillator strength of two-photon transition for the different hyperfine
Zeeman sublevels $(-2,-2),(-1,-1),(0,0),(1,1),(2,2)$ respectively. The energy
separation of the different sublevels is well explained by eqs.(\ref{Q-1}%
-\ref{Q-7}) when the bias magnetic field $B=220$ mG is applied. The magnetic
field is calibrated by the linear Zeeman shift of the hyperfine Zeeman
sublevels $(-2,-2),(2,2)$. For different magnetic field, we measured the
resonance frequency for the transitions $(-2,-2)$ and $(2,2)$, as shown in
Fig.\textit{\ref{f-3}}. After a linear fit, the slope is the magnetic field
intensity controlled by the current of the coils in the Raman interaction
area. The scaled method is similar to that of the quadratic Zeeman shift
measurement introduced in our paper. The scaled parameters come from earlier
references \cite{Steck,Bender1958a,Penselin1962a,Arimondo1977a}. The scale
factor of the magnetic field is $1576.9\pm1.3$ mG/A after the averaged measurements.%

\begin{figure}
[ptbh]
\begin{center}
\includegraphics[
trim=1.144065in 3.191197in 1.539595in 3.770597in,
height=2.1975in,
width=2.7432in
]%
{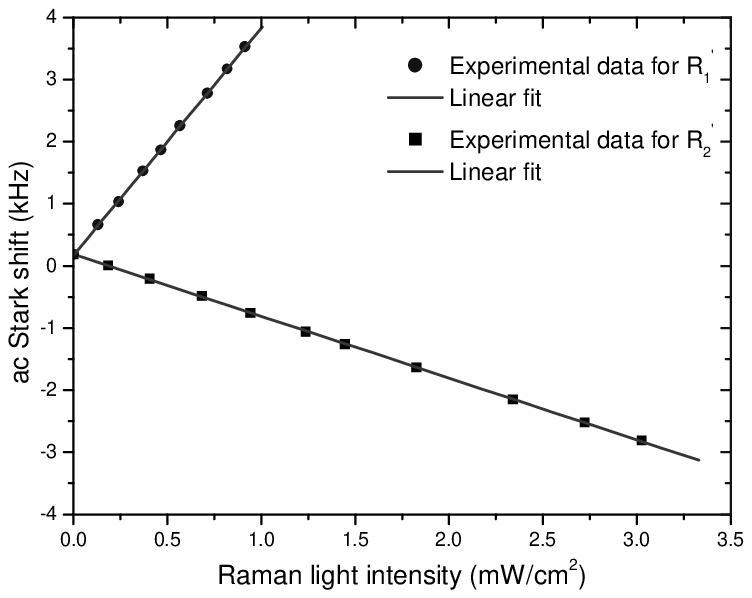}%
\caption{The differential ac Stark shift caused by imbalanced Raman beams
versus the Raman light intensity. The dots are the frequency shift induced by
$R_{1}^{^{\prime}}$\ while the squares are the frequency shift induced by
$R_{2}^{^{\prime}}$, where they are fitted linearly and the slopes are $3.66$
kHz/(mW/cm$^{2}$) and $-0.99$ kHz/(mW/cm$^{2}$) respectively. The ac Stark
shift can be cancelled by adjusting the intensity ratio to $1:3.67$ for the
one-photon detuning $\Delta=1.5$GHz.}%
\label{f-4}%
\end{center}
\end{figure}

The differential ac Stark shift caused by the imbalanced Raman beams will
induce a measurement noise in the determination of the quadratic Zeeman shift.
The difference between the ac Stark shifts of two hyperfine sublevels,
$\delta^{AC}=\Omega_{F=3,m_{F}=0}^{AC}-\Omega_{F=2,m_{F}=0}^{AC}$, can be
cancelled by optimizing the ratio of two Raman beams \cite{weiss1994a}. We
measure the frequency shift that is induced by one of the Raman beams
separately. In the experiment, we use two pairs of Raman beams (R$_{1}$,
R$_{2}$) and (R$_{1}^{^{\prime}}$, R$_{2}^{^{\prime}}$) along the magnetic
field $B$ and $B_{0}$, where $B$ and $B_{0}$ are $250$ mG and $100$ mG
respectively. The Raman beams (R$_{1}$, R$_{2}$) are used to measure the ac
Stark shift induced by the other Raman beams (R$_{1}^{^{\prime}}$,
R$_{2}^{^{\prime}}$). We carefully optimize the intensities of the Raman beams
($R_{1}$, $R_{2}$) along the magnetic field $B$\ to obtain a $\pi$-pulse. We
scan the frequency difference of the Raman beams (R$_{1}$, R$_{2}$), and the
resonant frequency of the hyperfine Zeeman sublevels ($0,0$) can be obtained
by a Gaussian fit for the different Raman light intensities (R$_{1}^{^{\prime
}}$, or R$_{2}^{^{\prime}}$). The detailed proceedure is similar to test of
quadratic Zeeman shift measurement in the paper. In the experiment, the Raman
beams (R$_{1}$, R$_{2}$) and (R$_{1}^{^{\prime}}$, R$_{2}^{^{\prime}}$) are
guided using single mode polarization maintained fiber. The intensity
instability is below one part in one thousand for each of the Raman beams. The
dots are the frequency shift that is induced by R$_{1}^{^{\prime}}$, while the
squares are the frequency shift that is induced by R$_{2}^{^{\prime}}$ in
Fig.\textit{\ref{f-4}}, where they are fitted linearly. The slopes are $3.66$
kHz/(mW/cm$^{2}$) and $-0.99$ kHz/(mW/cm$^{2}$) for R$_{1}^{^{\prime}}$, and
R$_{2}^{^{\prime}}$ respectively. The frequency shifts, induced by the
different Raman beams (R$_{1}^{^{\prime}}$, or R$_{2}^{^{\prime}}$), are
referenced to the separation of hyperfine sublevels ($3$ $035$ $732$ $436$)
\cite{Penselin1962a}. The non null values are mainly caused by the quadratic
Zeeman shift when the Raman beams (R$_{1}^{^{\prime}}$, R$_{2}^{^{\prime}}$)
are not applied in Fig.\textit{\ref{f-4}}. The ratio ($1:3.67$) of the two
slopes determines the cancellation of the ac Stark shift when the one-photon
detuning is $1.5$ GHz in our experiment. Therefore, we can cancel the ac Stark
shift by adjusting the ratio of two Raman beam intensities.%

\begin{figure}
[ptbh]
\begin{center}
\includegraphics[
trim=1.176157in 2.874547in 1.569279in 4.191674in,
height=2.1655in,
width=2.738in
]%
{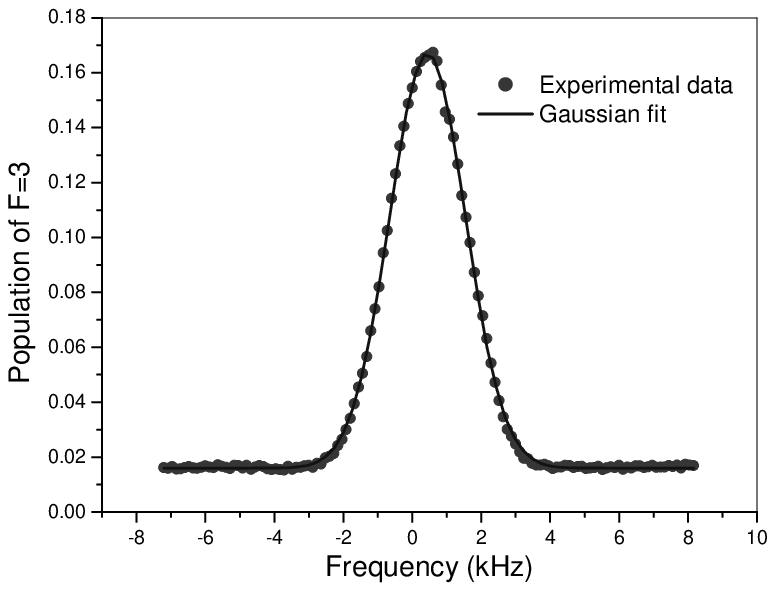}%
\caption{Frequency dependence of transition amplitude for a square Raman $\pi
$-pulse (R$_{1}$, R$_{2}$). The frequency is referenced to the hyperfine
separation of the two ground states. The dots are the experimental data with
$B=600$ mG, B$_{0}$=$0$, R$_{1}^{^{\prime}}$=$0$, R$_{2}^{^{\prime}}$=$0$.
while the solid line is the Gaussian fit. }%
\label{f-5}%
\end{center}
\end{figure}

\begin{figure}
[ptbh]
\begin{center}
\includegraphics[
trim=1.282059in 3.136176in 1.545211in 4.081632in,
height=2.1197in,
width=2.7371in
]%
{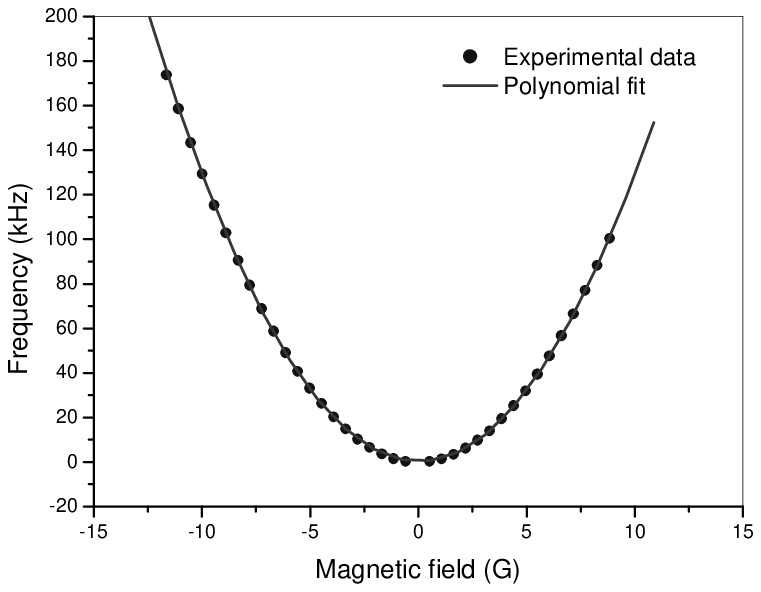}%
\caption{ Dependence of resonance frequency on the magnetic field intensity.
The dots are the frequency shift in the different magnetic field that is
obtained from Fig. \textit{\ref{f-5}}, while the line is the experimental fit
by a polynomial function.}%
\label{f-6}%
\end{center}
\end{figure}

After the magnetic field compensation and the cancellation of the ac Stark
shift, their influence is considerably decreased in the measurement of the
quadratic Zeeman shift. The Raman beams are generated by an acousto-optical
modulator(Brimrose, $1.5$\ GHz) driven by microwave generator (Agilent
$8257$C) which is locked by a H-maser. The arrangement of the Raman laser is
similar to our previous work \cite{Wang2007a}. We carefully optimize the
intensities of the Raman beams ($R_{1}$, $R_{2}$) along the magnetic field
$B$\ to obtain a $\pi$-pulse, where $B_{0}$, $R_{1}^{^{\prime}}$ and
$R_{2}^{^{\prime}}$ are not used. The instability of the ratio of the Raman
beams ($R_{1}:R_{2}=1:3.67$) is below $10^{-5}$ in the experiment. We scan the
frequency difference of the Raman beams (R$_{1}$, R$_{2}$), and observe a
typical stimulated Raman transition which shows the population versus
frequency difference between the two Raman beams in Fig.\textit{\ref{f-5}} at
a magnetic field $B=600$ mG, where the frequency is referenced to the
separation between the two ground states ($3$ $035$ $732$ $436$ Hz)
\cite{Penselin1962a}. In our experiment, the intensity profile of the Raman
beams is a Gaussian distribution and the line width is mainly limited by the
transition time because the spontaneous can be ignored in large one-photon
detuning. In such case, the population dependence on the two-photon detuning
is a Gaussian profile \cite{Demtroder2003a}. The central frequency is obtained
from a Gaussian fit. We have made a series of such curves for different
magnetic fields, and the dependence of the frequency shift on the magnetic
field is shown in Fig.\textit{\ref{f-6}}. The frequency shift depends on the
magnetic field and it is fitted by a polynomial function (The maximum power is
$2$), while the quadratic dependence is for the quadratic Zeeman shift. We
measured a series of values as shown in table $%
1%
$, and the average frequency shift induced by the quadratic Zeeman effect for
the hyperfine Zeeman sublevels ($5S_{1/2},F=2,m_{F}=0\rightarrow
5S_{1/2},F=3,m_{F}=0$) is $1296.8$ Hz/G$^{2}$. The measurement uncertainty
comes mainly from the calibrated magnetic field and the fitted error. As shown
in table $%
1%
$, the averaged uncertainty of the quadratic Zeeman shift is $2.1$ Hz/G$^{2}$
and $2.5$\ Hz/G$^{2}$ for the scaled magnetic field and the fitted error
respectively. The final result for the quadratic Zeeman shift is
$1296.8\pm3.3$\ Hz/G$^{2}$ by using an independent error source model, which
is in good agreement with the calculation result \cite{Steck} within our
measurement precision. The result shows that the second perturbation theory is
sufficient when the magnetic field is less than $1$ mT \cite{Itano2000a}. The
ac Stark shifts induce a systematic shift of the ground-state hyperfine
splitting. This does not influence the value of the quadratic Zeeman shift
when a quadratic dependence term of the polynomial function is chosen as shown
in Fig.\textit{\ref{f-6}}. The fitted error, which is induced by the
instability of the Raman beams, is decreased when the cancellation ratio of
the Raman beams ($1:3.67$) is applied in the experiment.%

In the atom interferometer, the bias magnetic field is applied through the
interference area. Although the atoms are always kept in magnetically
insensitive states with $m_{F}=0$, these states still show a quadratic Zeeman
shift that induces a relative frequency shift of two ground states. This
effect is big enough to require well controlled magnetic fields and extensive
magnetic field shielding to achieve the millihertz frequency stability
necessary for gravity measurements at the $1\mu$G level \cite{Peters}. For the
rotation rate measurement, the quadratic Zeeman shift should be known
accurately when considering the accuracy necessary to determine the rotation
rate of the earth. The sensitivity of the rotation signal to the various bias
magnetic field was determined in detailly performed in the dual atomic
interferometer gyroscope, and the bias magnetic field caused a phase shift
$2\times10^{-6}\Omega_{E}$/mG for the rotation measurement in the system
\cite{Gustavson}, which is mainly induced by the quadratic Zeeman shift. In
our experiment, the precision of the quadratic Zeeman shift is mainly limited
by the measurement time, and it can be measured even more accurately by
decreasing the atomic flight velocity and increasing the Raman beam diameter,
and by using the separated oscillation field method in a weak magnetic field
\cite{Bize1999a}. However, our result provides helpful data for higher
precision measurement of the quadratic Zeeman shift of $^{85}$Rb, even for the
accuracy of the rotation rate measurement of the atom-interferometer gyroscope.

Table $%
1%
$ Experimental data for the determination of the quadratic Zeeman shift of
hyperfine sublevels ($5S_{1/2},F=2,m_{F}=0\rightarrow$ $5S_{1/2},F=3,m_{F}=0$)
of $^{85}$Rb.

\ \
\begin{tabular}
[c]{cccc}\hline\hline
$Run$ & $%
\begin{array}
[c]{c}%
Frequency\\
shift\\
\text{(Hz/G}^{2}\text{)}%
\end{array}
$ & $%
\begin{array}
[c]{c}%
Scaled\\
error\\
\text{ (Hz/G}^{2}\text{)}%
\end{array}
$ & $%
\begin{array}
[c]{c}%
Fitted\\
error\\
\text{(Hz/G}^{2}\text{)}%
\end{array}
$\\\hline
$1$ & $1294.2$ & $2.1$ & $2.9$\\
$2$ & $1294.1$ & $2.1$ & $2.9$\\
$3$ & $1295.7$ & $2.1$ & $2.3$\\
$4$ & $1296.1$ & $2.1$ & $2.2$\\
$5$ & $1298.6$ & $2.1$ & $1.9$\\
$6$ & $1298.7$ & $2.1$ & $1.9$\\
$7$ & $1298.7$ & $2.1$ & $1.9$\\
$8$ & $1298.6$ & $2.1$ & $1.9$\\
$Average$ & $1296.8$ & $2.1$ & $2.5$\\
$Total$ & \multicolumn{3}{c}{$1296.8\pm3.3$ (Hz/G$^{2}$)}\\\hline\hline
\end{tabular}

\begin{flushleft}
\textbf{5. Conclusion}
\end{flushleft}

In summary, we analyzed the energy of the hyperfine sublevels of two ground
states of $^{85}$Rb in the magnetic field. We demonstrated experimentally the
coherent population transfer of the hyperfine sublevels between two ground
states by the stimulated Raman transition. The ac Stark shift was
experimentally studied by measuring the ac Stark frequency shift dependence on
the Raman beam intensity, and it was cancelled by adjusting the ratio of two
Raman beam intensities. We measured the quadratic Zeeman shift of the ground
states using the coherent population transfer by a stimulated Raman
transition. The error analysis shows that the quadratic Zeeman shift was
measured to Hz level for magnetically insensitive states $5S_{1/2}%
,F=2,m_{F}=0\rightarrow$ $5S_{1/2},F=3,m_{F}=0$ in the experiment. This result
provides helpful data to improve the accuracy of the atom-interferometer
gyroscope in future.

\begin{flushleft}
\textbf{Acknowledgments}
\end{flushleft}

We acknowledge the financial support from the National Basic Research Program
of China under Grant Nos. 2005CB724505, 2006CB921203, and from the National
Natural Science Foundation of China under Grant No.10774160. We thank
Professor J. P. Connerade and Professor L. You for useful comment and discussion.

\end{document}